\documentclass[aps,prl,twocolumn,superscriptaddress,nobibnotes,longbibliography,10pt]{revtex4-2}
\usepackage{graphicx}
\usepackage{dcolumn}
\usepackage{bm}
\usepackage{makecell}
\usepackage[utf8]{inputenc}
\usepackage[T1]{fontenc}
\usepackage{mathptmx} 
\usepackage{etoolbox}
\usepackage{xcolor}
\usepackage{float}
\usepackage{amsmath}
\usepackage{amssymb}
\usepackage{hyperref}
\usepackage{comment}
\usepackage[normalem]{ulem} 
\usepackage{amsmath}
\usepackage{bm}

\begin{document}


\title{Transition Dipole Rotation Beyond the Condon Approximation in Single hBN Quantum Emitters}

\author{Serkan Pa\c{c}al}
\affiliation{Department of Physics, \.{I}zmir Institute of Technology, \.{I}zmir, 35430, Turkey}

\author{Chanaprom Cholsuk}
\affiliation{Department of Computer Engineering, TUM School of Computation, Information and Technology, Technical University of Munich, 80333 Munich, Germany}
\affiliation{Munich Center for Quantum Science and Technology (MCQST), 80799 Munich, Germany}

\author{Mouli Hazra}
\affiliation{Department of Computer Engineering, TUM School of Computation, Information and Technology, Technical University of Munich, 80333 Munich, Germany}
\affiliation{Munich Center for Quantum Science and Technology (MCQST), 80799 Munich, Germany}

\author{\c{C}a\u{g}lar Samaner}
\affiliation{Department of Physics, \.{I}zmir Institute of Technology, \.{I}zmir, 35430, Turkey}

\author{\"Ozg\"ur \c{C}ak\i r}
\affiliation{Department of Physics, \.{I}zmir Institute of Technology, \.{I}zmir, 35430, Turkey}

\author{Tobias Vogl}
\affiliation{Department of Computer Engineering, TUM School of Computation, Information and Technology, Technical University of Munich, 80333 Munich, Germany}
\affiliation{Munich Center for Quantum Science and Technology (MCQST), 80799 Munich, Germany}

\author{Serkan Ate\c{s}}
\email{serkan.ates@sabanciuniv.edu}
\affiliation{Faculty of Engineering and Natural Sciences, Sabanc\i\ University, \.{I}stanbul, 34956, Turkey}

\begin{abstract}
The design of polarization-encoded quantum interfaces relies on the assumption that solid-state emitters possess static transition dipoles defined by the host lattice symmetry. Here, we demonstrate that the transition dipole moment of single hexagonal boron nitride quantum emitters is not a static property but rotates as a function of photon energy. Through high-resolution energy-resolved spectroscopy, we reveal a continuous rotation of the emission dipole orientation reaching up to $40^{\circ}$ across the vibronic manifold at room temperature, driven by coupling to the phonon bath. This spectral rotation is effectively suppressed at cryogenic temperatures (6~K), where the acoustic phonon population is negligible, identifying thermally activated lattice vibrations as the primary driver of the reorientation. First-principles calculations on two representative defects spanning weak and strong electron-phonon coupling regimes confirm that phonon-displaced geometries produce a systematic deviation of the transition dipole orientation from the zero-phonon line, with the magnitude scaling with vibronic coupling strength. The experimental observations and calculations demonstrate that single quantum emitters can operate beyond the Condon approximation, with the transition dipole acquiring a dependence on the instantaneous nuclear configuration. Our results identify a fundamental limit for polarization fidelity in solid-state quantum networks and connect solid-state single-emitter physics to a class of effects previously accessible only in ensemble measurements in molecular and biological spectroscopy.
\end{abstract}

\maketitle 

Two-dimensional (2D) materials have opened exciting opportunities in quantum photonics, enabling the realization of atomically thin platforms with robust, room-temperature single-photon sources \cite{Kianinia.Aharonovich.2022, Esmann.Anton-Solanas.2024}. Among these materials, hexagonal boron nitride (hBN) has gained particular attention due to its wide bandgap ($\sim 6$~eV), excellent chemical stability, and the presence of optically active point defects that emit bright, spectrally narrow, and linearly polarized photons \cite{Tran2016, Nikolay2019, Dietrich2018, Cakan2025}. These properties make hBN-based emitters appealing candidates for applications such as quantum communication \cite{Samaner2022, Juboori2023, Tapsin.Ates.2025}, quantum sensing \cite{Rizzato2023, Sortino2024}, and quantum memories \cite{Nateeboon2024, cholsuk_identifying_2024}.

A defining feature of these defect-based emitters is the polarization of their optical transitions, which is central to quantum protocols where the polarization state encodes information. In hBN, the hexagonal lattice symmetry naturally supports linearly polarized dipole transitions~\cite{Jungwirth2017}, enabling highly indistinguishable photons under resonant excitation \cite{Gerard.Delteil.2026}. However, practical non-resonant excitation introduces significant dephasing and a loss of indistinguishability due to increased interaction with the phonon bath \cite{Fournier.Delteil.2023, Sontheimer.Benson.2017}. While the impact of phonons on spectral lineshapes and coherence is well-documented \cite{Exarhos2017, Khatri2019, Kubanek.Kubanek.2022, Martinez.Solanas.2026}, their influence on the orientation of the transition dipole itself remains unexplored. Most standard models in solid-state quantum optics rely on the Condon approximation, which assumes a static transition dipole moment determined solely by the crystal field symmetry \cite{Kumar2024, Jungwirth2017}, treating the dipole as a constant that is insensitive to nuclear motion or temperature. The breakdown of this approximation through vibronic coupling~\cite{Herzberg.Teller.1933} is well-documented in the spectroscopy of complex molecules, photosynthetic aggregates, and organic chromophores~\cite{Small.Small.1971}, where it governs spectral intensities and symmetry-forbidden transitions, manifesting as a scalar modification of the optical response. However, the vectorial rotation of the transition dipole orientation as a function of nuclear coordinate has not previously been resolved at the single-emitter level in any solid-state system. Recent theoretical work has suggested that coordinate-dependent transition dipole contributions can be significant even in electric-dipole allowed transitions of solid-state defects~\cite{Quan.Duan.2025}, yet a direct experimental demonstration at the single-emitter level has been missing. Despite recent advances in understanding the spectral and temporal polarization characteristics of hBN defect centers~\cite{Kumar2024, Samaner2025}, the vectorial consequence of vibronic coupling on the transition dipole orientation in atomically thin 2D materials has not been directly addressed.
\begin{figure*}[t!]
\includegraphics[width=0.85\textwidth]{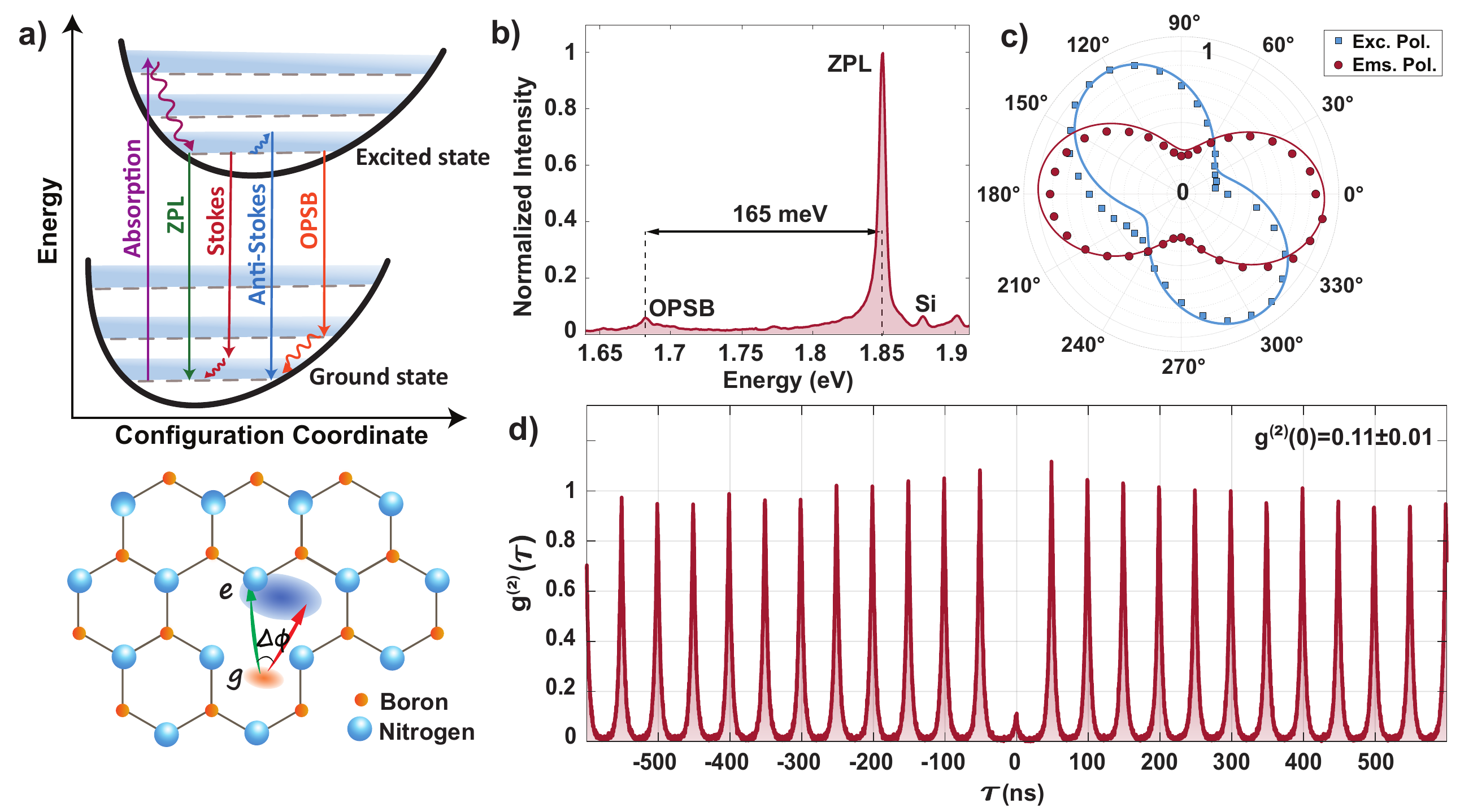}
    \caption{(a) (top) Energy level diagram illustrating the vibronic emission pathways of a localized two-level defect, including the zero-phonon line (ZPL), Stokes, anti-Stokes, and optical phonon sideband (OPSB) transitions, each potentially exhibiting a distinct transition dipole orientation. (bottom) Schematic of a defect center in hBN, showing a localized electronic state with a rotation of the emission dipole ($\Delta\phi$) driven by phonon-induced atomic displacements, illustrating the coordinate-dependent transition dipole demonstrated in this work. The atomic configuration is provided as an illustrative model to demonstrate the general vibronic mechanism, rather than an exact representation of the experimental defect. (b) Normalized PL spectrum at room temperature, highlighting a sharp ZPL at $1.848$~eV ($671$~nm) and the OPSB at $1.683$~eV ($736$~nm). (c) Polarization-resolved excitation and emission dipole orientations mapped in polar coordinates, showing an angular offset between the excitation (squares) and emission (circles) dipoles under non-resonant excitation. (d) Second-order autocorrelation function measured under pulsed excitation, yielding $g^{(2)}(0) = 0.11 \pm 0.01$, confirming single-photon emission.}
  \label{figure1}
\end{figure*}
In low-dimensional systems where strong electron-phonon coupling governs the spectral lineshape~\cite{Wigger2019, White2021}, the possibility of a coordinate-dependent transition dipole presents a fundamental limit to polarization stability. In this work, we demonstrate that vibronic lattice fluctuations in single hBN quantum emitters fundamentally modulate the transition dipole orientation as a function of photon energy, constituting a non-Condon effect in which the transition dipole moment depends on the instantaneous nuclear configuration rather than remaining fixed at its equilibrium value. Through high-resolution spectral mapping combined with full Stokes parameter analysis \cite{Schaefer2007}, we resolve this modulation as a continuous rotation of the emission dipole orientation across the vibronic manifold of single hBN defects. We perform these measurements at both room and cryogenic temperatures to isolate thermally activated phonon contributions and to eliminate spectral averaging from closely-spaced electronic transitions~\cite{Becher2019}. First-principles 0~K calculations reveal that optical phonon modes \cite{Wigger2019} produce a systematic deviation of the transition dipole orientation at phonon-displaced geometries, while the dominant contribution at room temperature, identified through the thermal suppression of the rotation at 6~K, arises from the acoustic phonon bath \cite{Hoese2020}. Taken together, these results show how phonon-induced atomic displacements perturb the electronic wavefunctions and drive the polarization orientation, establishing that the transition dipole in hBN is not a rigid property but an energy-dependent degree of freedom.

\section{Results and discussion}

To establish the physical basis for the phonon-induced breakdown of the Condon approximation, Figure~\ref{figure1}a shows the vibronic framework for a localized two-level defect. In the standard Franck-Condon picture, the transition dipole moment is coordinate-independent, dictating a constant polarization axis regardless of whether the emission originates purely from the zero-phonon line (ZPL) or involves vibrational manifolds. However, in the two-dimensional hBN matrix, coupling to the continuum of acoustic phonon states (blue shaded regions) and discrete optical phonon modes (dashed lines) can modify the electronic wavefunction and thereby alter the transition dipole orientation. As illustrated in the top panel of Figure~\ref{figure1}a, transitions involving distinct vibrational pathways can therefore exhibit different transition dipole orientations. These pathways include the isolated ZPL (green), Stokes (red), and anti-Stokes (blue) emissions. The lower panel of Figure~\ref{figure1}a presents an illustrative model of this mechanism for a localized electronic state. While not an exact atomic representation of the experimental defect, this schematic framework demonstrates how local structural perturbations physically reorient the emission dipole within the lattice, resulting in a distinct angular separation $\Delta\phi$ between optical transitions.
\begin{figure*}[t!]
  \includegraphics[width=0.8\textwidth]{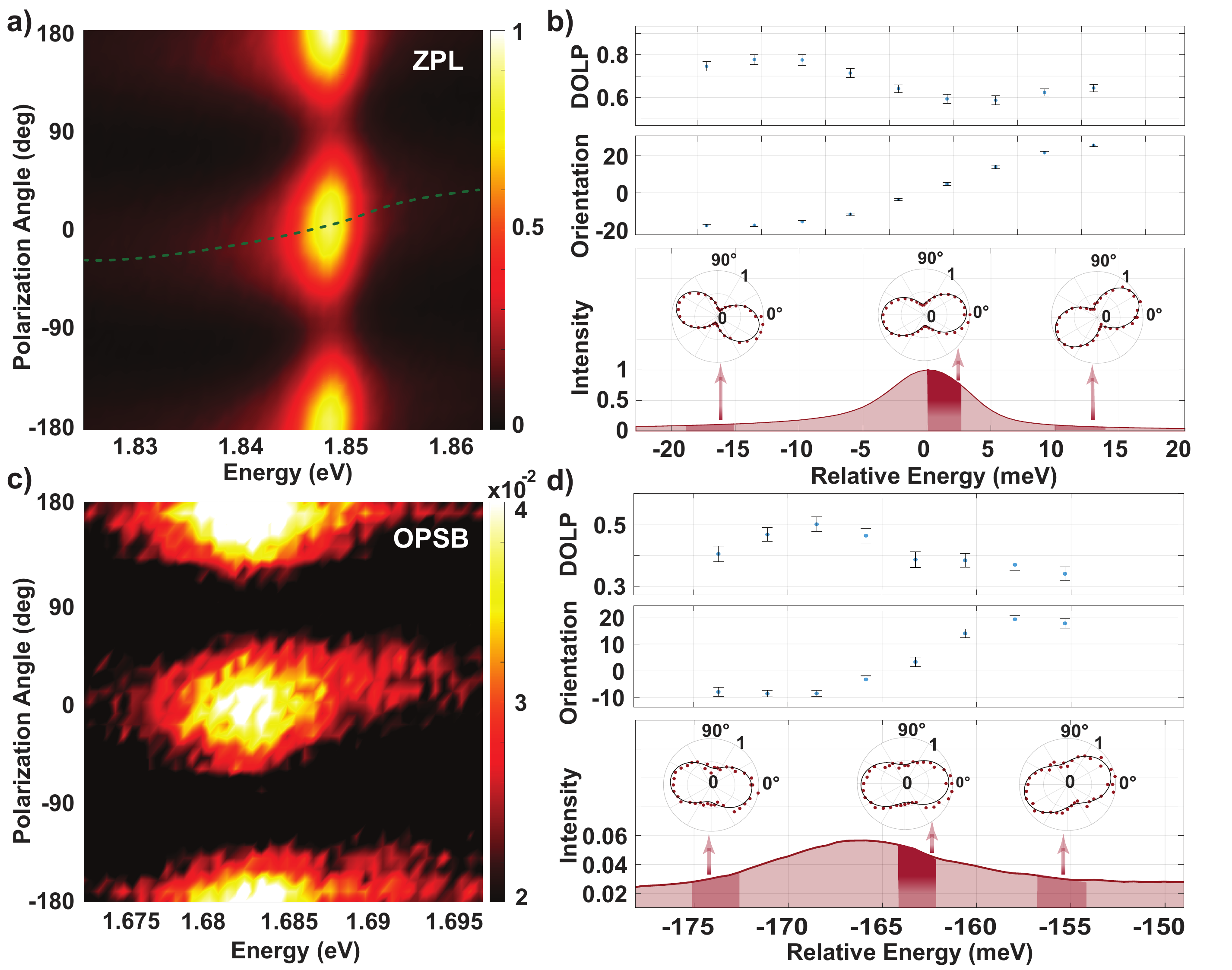}
\caption{Energy-resolved linear polarization analysis of the hBN defect center. (a,~c) Emission intensity maps plotted as a function of photon energy and analyzer angle, where the systematic shift of the peak emission axis with photon energy reveals the continuous rotation of the transition dipole across the ZPL and OPSB regions, respectively. (b,~d) Quantitative polarization parameters extracted from $\sim$4~meV spectral slices. Each panel displays: (top) the degree of linear polarization (DOLP), demonstrating a consistently high directional dipole character across the emission band; (middle) the polarization orientation angle, revealing a continuous sweep exceeding $40^{\circ}$ and providing direct evidence for an energy-dependent transition dipole; and (bottom) the emission spectrum normalized to the peak ZPL intensity shown in panel~(b), with representative polar plots at the indicated spectral positions. Panel~(d) shows the equivalent analysis for the OPSB spectral window ($155$--$175$~meV below the ZPL). The energy axis is referenced relative to the ZPL center ($E - E_\mathrm{ZPL}$) to allow direct comparison of the distinct electron-phonon coupling regimes.}
  \label{figure2}
\end{figure*}

Figure~\ref{figure1}b presents the room-temperature photoluminescence (PL) spectrum of an isolated hBN defect under pulsed $637$~nm ($1.946$~eV) laser excitation. The emission is dominated by a sharp ZPL at $1.848$~eV ($671$~nm) and a distinct optical phonon sideband (OPSB) at $1.683$~eV ($736$~nm), corresponding to the characteristic $\sim165$~meV high-energy Raman-active optical phonon mode of the hBN lattice~\cite{Khatri2019}. Asymmetric spectral broadening of the ZPL indicates electron-phonon coupling to the continuous acoustic manifold (APSBs)~\cite{Ari.Ates2025}. Spectrally integrated polarization measurements, mapped in polar coordinates (Figure~\ref{figure1}c), characterize both the absorption and emission dipole orientations. Modulating the incident laser polarization yields an excitation polarization visibility of $0.65$ for the ZPL intensity. Aligning the excitation laser to this optimal axis yields a degree of linear polarization (DOLP) of $0.71$ for the ZPL emission, filtered over a $28~\text{meV}$ spectral window. The finite angular offset between excitation and emission dipoles is typically attributed to non-resonant excitation conditions~\cite{Jungwirth2017}, though the energy-resolved analysis presented below suggests it may also reflect the photon-energy dependence of the transition dipole orientation demonstrated in this work. Crucially, pulsed second-order autocorrelation measurements yield an antibunching dip of $g^{(2)}(0) = 0.11 \pm 0.01$ (Figure~\ref{figure1}d). This high single-photon purity rules out the possibility of overlapping optical transitions from spatially distinct defects, ensuring that the filtered ZPL emission broadened by acoustic phonons is attributed to a single quantum emitter.

While the spectrally integrated linear polarization measurements in Figure~\ref{figure1}c confirm the highly polarized nature of the defect, they implicitly assume a fixed transition dipole and average over the energy-dependent emission. To determine whether the polarization state is structurally locked or dynamically coupled to the vibrational manifold, we performed high-resolution, spectrally resolved linear polarization analysis. The resulting emission intensity map (Figure~\ref{figure2}a) visualizes the intensity as a function of emission energy and polarization angle, where the green dashed trace tracks a systematic rotation of the peak emission axis. 

To quantitatively evaluate this effect, the spectral map was discretized into $\sim$4~meV bins, and each angular emission profile was fitted with a Malus' law function of the form $I(\theta) = I_{\max}\cos^2(\theta - \theta_0) + I_{\min}$ to extract the orientation angle $\theta_0$ and the degree of linear polarization $\text{DOLP} = (I_{\max} - I_{\min})/(I_{\max} + I_{\min})$. As shown in Figure~\ref{figure2}b (top), the DOLP remains consistently high (0.6 to 0.8) throughout the emission band. This confirms that the photon emission maintains a highly directional dipole character across all energies, ruling out random depolarization artifacts. The combination of high DOLP and large orientation sweep is the distinct experimental signature of a vibronically modulated transition dipole: the dipole remains well-defined at every photon energy, only its direction changes. However, the orientation angle (Figure~\ref{figure2}b, middle) exhibits a continuous, large-amplitude spectral sweep. The transition dipole smoothly rotates as a function of photon energy, providing direct experimental evidence for an energy-dependent transition dipole. The angular deviation appears larger on the anti-Stokes side of the ZPL, consistent with a stronger perturbation of the excited-state electronic wavefunction by acoustic lattice distortions. Anti-Stokes emission originates from thermally populated higher vibrational levels of the excited electronic state, which sample a broader range of nuclear configurations and therefore probe a larger portion of the dipole's dependence on nuclear coordinate. While the reduced signal intensity on the anti-Stokes side limits precise quantification of this asymmetry from a single emitter, the directional trend is physically consistent with stronger vibronic coupling in the excited-state vibrational manifold compared to the ground state. Representative polar plots (Figure~\ref{figure2}b, bottom) explicitly visualize this energy-dependent angular divergence relative to the ZPL center. A spectrally resolved characterization of the full optical path using a polarized broadband source confirms that the detection system introduces no measurable energy-dependent polarization rotation, with the orientation angle stable within $\pm2^{\circ}$ and DOLP $>0.95$ across the full spectral window (Supplementary Section S2). The following energy-resolved analysis, therefore, directly reflects the intrinsic polarization properties of the emitter.

This nuclear-configuration dependence of the dipole is not limited to the low-energy acoustic manifold of the ZPL. As shown in Figure~\ref{figure2}c,d, the OPSB,  located 165~meV below the ZPL and arising from the characteristic Raman-active optical mode of the hBN lattice, independently reproduces the same behavior in a spectrally remote emission channel. Two distinct manifestations are observed. First, the center of the OPSB displays a $5^{\circ}$ discrete orientation offset relative to the ZPL, constituting a direct inter-band polarization shift between two emission channels of the same single emitter. Second, a continuous $30^{\circ}$ intra-band rotation is observed across the OPSB spectral width (155--175~meV), arising from simultaneous coupling to acoustic phonons that broaden the optical sideband~\cite{Krummheuer2002}. The internal consistency across these two phononic regimes is significant. Any instrumental artifact or sample inhomogeneity capable of producing the $40^{\circ}$ rotation near the ZPL would need to independently reproduce a qualitatively identical pattern in a spectrally separate window through an entirely different emission mechanism. This consistency therefore confirms that the vectorial rotation is an intrinsic, unified feature of the emitter’s non-Condon dipole response.

\begin{figure*}[t!]
  \includegraphics[width=0.8\textwidth]{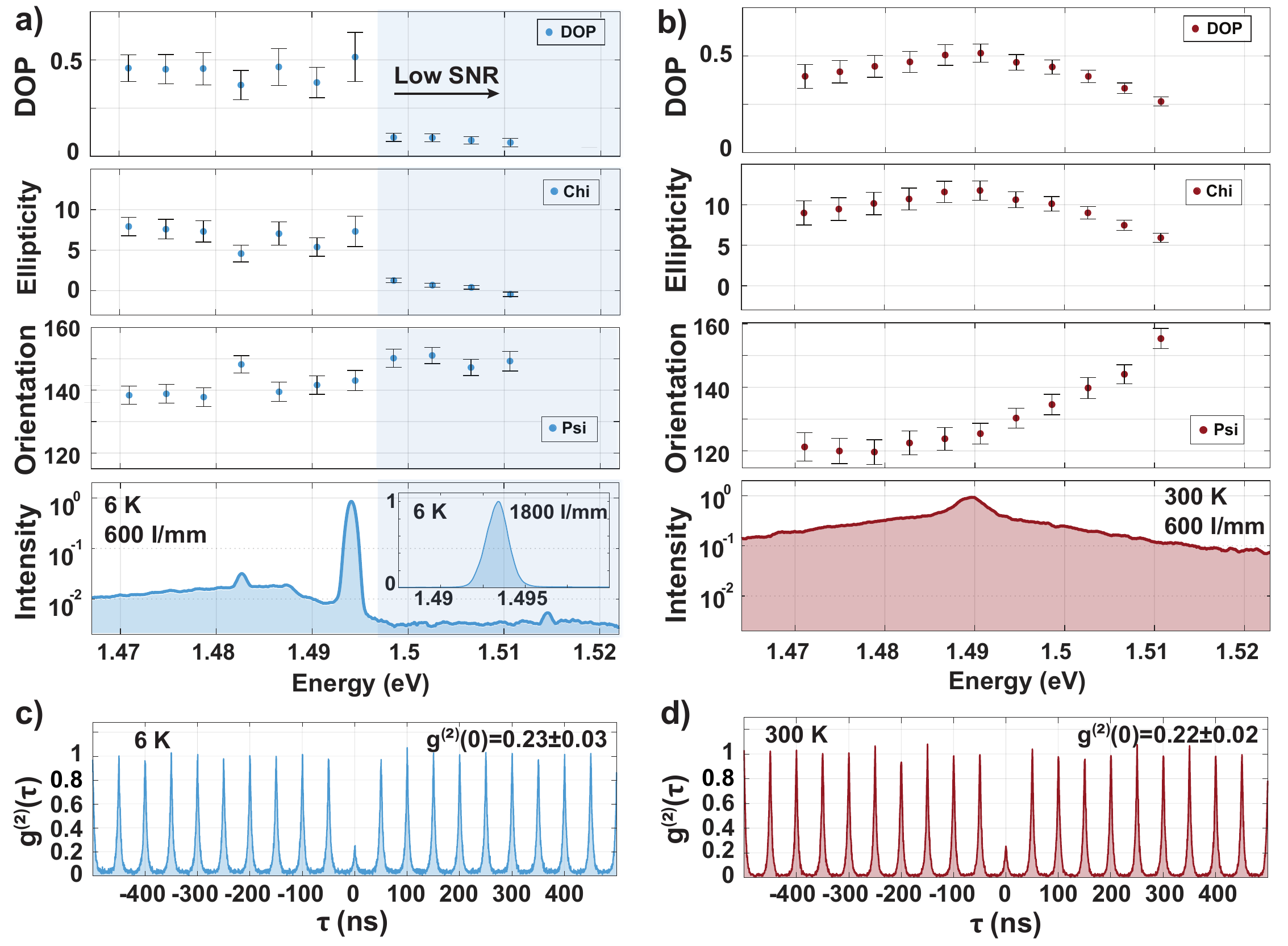}  
  \caption{Thermal suppression of the coordinate-dependent dipole rotation and single-photon verification across the thermal cycle. (a,~b) Energy-resolved Stokes polarimetry and corresponding PL spectra at $6$~K and $300$~K, respectively. Each panel displays a vertical stack of (top to bottom): the degree of polarization 
(DOP), ellipticity ($\chi$), orientation angle ($\psi$) in degrees, and emission intensity. The $40^{\circ}$ orientation sweep observed at $300$~K is effectively suppressed at $6$~K, where the acoustic phonon population is negligible, identifying thermally activated lattice vibrations as the driver of the room-temperature reorientation. The negligible ellipticity ($\chi$) confirms that the emission remains linearly polarized across both thermal regimes. Shaded blue areas indicate regions of low signal-to-noise ratio (SNR). (c,~d) Second-order autocorrelation functions $g^{(2)}(\tau)$ measured at $6$~K and $300$~K under identical pulsed excitation. The consistent antibunching values ($g^{(2)}(0) \approx 0.22$) confirm the single-photon identity of the emitter across the thermal cycle.}
  \label{figure3}
\end{figure*}

To provide a definitive test of the phonon-driven reorientation mechanism, we perform energy-resolved Stokes polarimetry at 6~K, the thermal regime in which the acoustic phonon population is negligible and the transition dipole is predicted to recover its static, equilibrium-geometry orientation. For this validation, we utilized a second representative defect (Emitter 2), which exhibits a single, isolated optical transition at low temperatures (Figure~\ref{figure3}a, bottom panel). This ensures that the polarization dynamics remain distinct and are not obscured by the spectral overlap inherent to room-temperature hBN ensembles or the presence of independent, closely-spaced electronic transitions within the same defect site~\cite{Becher2019}. Building on established time-resolved Stokes polarimetry~\cite{Samaner2025}, we utilize the rotating quarter-wave plate (RQWP) method~\cite{Schaefer2007} to resolve the full Stokes vector in the energy domain. This approach enables the simultaneous determination of the polarization orientation ($\psi$) and ellipticity ($\chi$), providing a comprehensive description of the vibronic state across both thermal regimes.

Figure~\ref{figure3}a,b summarizes the contrast between cryogenic and room-temperature measurements for Emitter 2. At 6~K (Figure~\ref{figure3}a), the defect exhibits a sharp, resolution-limited ZPL at 1.494~eV (829~nm). In this cryogenic limit, where the acoustic phonon population is negligible~\cite{Grosso2020}, the orientation angle $\psi$ remains spectrally flat across the emission line and the degree of polarization (DOP) remains consistently high, while the anti-Stokes signal is naturally suppressed. We note that while $\chi$ approaches zero in the low-signal anti-Stokes region, this reflects the greater susceptibility of the circular Stokes component to the noise floor compared to the linear components from which $\psi$ is derived~\cite{Kozlov2018}, rather than a physical change in the polarization state. The 6~K data therefore confirms that the transition dipole is locked in its equilibrium configuration when thermal phonon excitation is absent, identifying phonon population as the direct driver of the room-temperature rotation.

In contrast, the $300$~K data in Figure~\ref{figure3}b reveals a pronounced broadening of the emission spectrum due to enhanced phonon interactions~\cite{Ari.Ates2025}. This spectral broadening is accompanied by a strong spectral dependence of the polarization parameters. While the DOP remains relatively high, confirming a well-defined dipole character, the orientation angle $\psi$ exhibits a smooth, continuous rotation of approximately $40^{\circ}$ across the emission band. This behavior identifies the dipole reorientation as a thermally activated process. The stability of $\psi$ at $6$~K compared to the $40^{\circ}$ sweep at $300$~K provides definitive evidence that the observed polarization rotation is a dynamic, phonon-induced effect rather than a static structural property of the defect.

The pronounced thermal evolution of the emission environment, visualized in the lower panels of Figure~\ref{figure3}a,b, directly correlates with the onset of the dipole rotation. While the cryogenic spectrum is dominated by a resolution-limited ZPL, the room-temperature manifold exhibits the broad, phonon-assisted profile characteristic of intensified electron--phonon coupling~\cite{Jungwirth2016, Cusco2016}. This transformation confirms that the observed rotation is not a secondary spectral effect but a fundamental modification of the transition dipole driven by the thermal activation of lattice modes. We note that the high-energy OPSB was omitted for Emitter 2 across both temperature regimes. This is due to the inherent efficiency limits of the optical setup in the spectral range above $900$~nm, where a combination of diminished detector quantum efficiency and optical throughput results in a prohibitive signal-to-noise ratio.

The single-photon nature of the emission from Emitter 2 is also confirmed by the second-order autocorrelation functions in Figure~\ref{figure3}c,d. Under pulsed excitation, the defect maintains high single-photon purity with $g^{(2)}(0) = 0.23 \pm 0.03$ at $6$~K and $0.22 \pm 0.02$ at $300$~K. To ensure maximum collection efficiency across the broad, room-temperature vibronic manifold, these measurements utilized an $800$~nm long-pass filter. The slightly elevated $g^{(2)}(0)$ values result from the inclusion of the long-wavelength background captured by this wide-band filtering, rather than a loss of emitter purity, as evidenced by the consistent antibunching value maintained throughout the thermal cycle.

Additional data in the Supplementary Information support the intrinsic nature of this effect. Measurements across multiple defect centers reveal that the magnitude of the rotation varies between emitters, consistent with the sensitivity of the dipole orientation to the local nuclear environment and electron-phonon coupling strength~\cite{Jha2022}. Emitters exhibiting negligible spectral rotation are consistent with a weaker vibronic coupling regime, analogous to the weak-coupling defect in the first-principles calculations presented below.

\section{Polarization Analysis with Density Functional Theory}
\begin{figure*}[t!]
    \centering
    \includegraphics[width=0.85\textwidth]{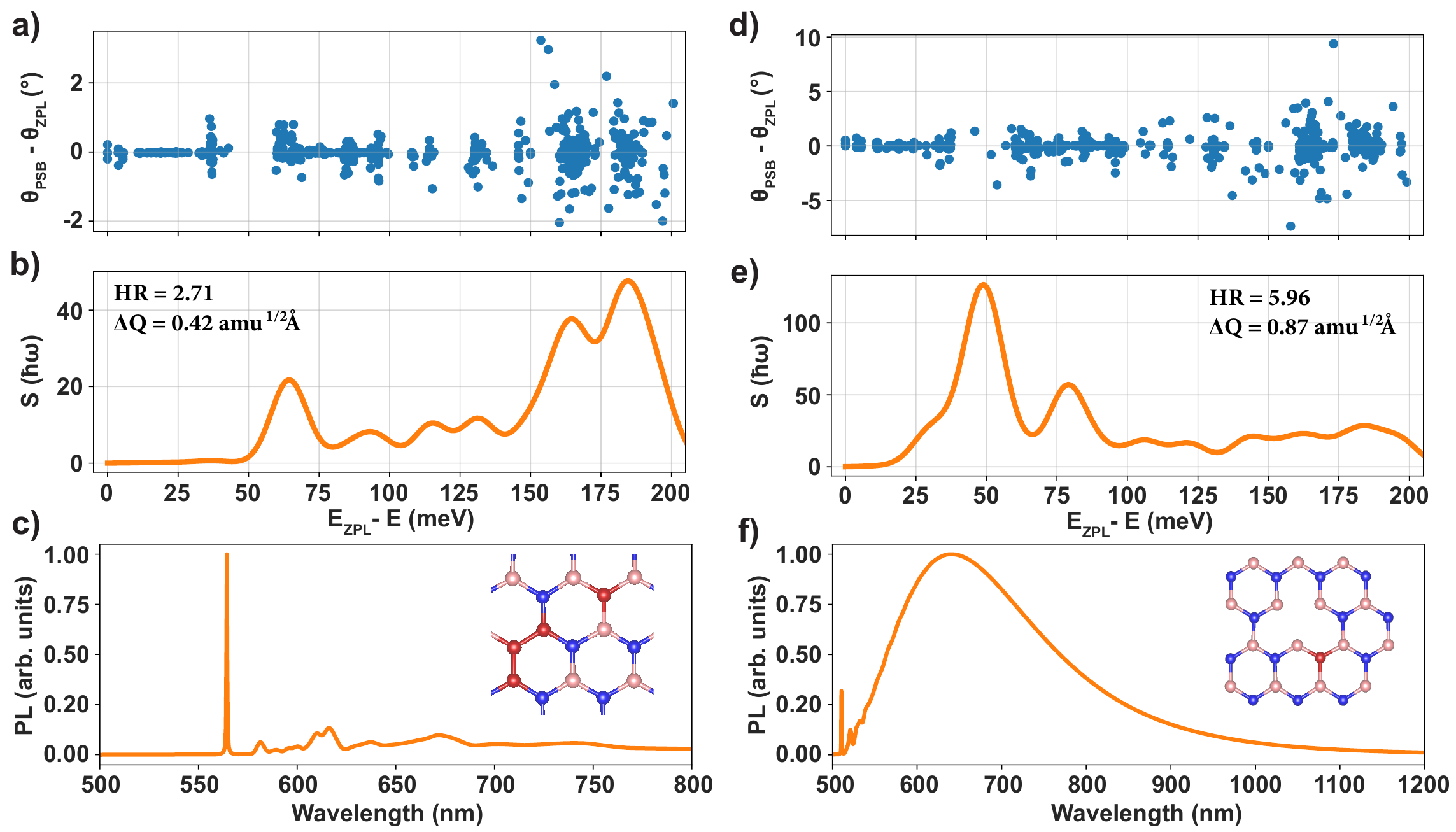}
    \caption{First-principles polarization rotation in the phonon sideband for defects spanning weak and strong vibronic coupling regimes in bulk hBN. (a--c) Results for the $C_{\text{B}}C_{\text{N}}C_{\text{B}}C_{\text{N}}$ defect (weak coupling, HR~$= 2.71$) and (d--f) the $C_{\text{N}}V_{\text{N}}$ defect (strong coupling, HR~$= 5.96$). (a,~d) Polarization orientation difference $\theta_\mathrm{PSB} - \theta_\mathrm{ZPL}$ resolved by individual optical phonon modes, showing maximum deviations of $2.7^{\circ}$ and $\sim10^{\circ}$ respectively, demonstrating that rotation magnitude scales with vibronic coupling strength. (b,~e) Calculated PSB spectral functions, with the Huang-Rhys factor HR and total configuration coordinate displacement $\Delta Q$ indicated for each defect. (c,~f) Resulting photoluminescence spectra showing the ZPL and phonon sideband distribution. The atomic models illustrate the defective layer environment, with pink, blue, and red spheres denoting boron, nitrogen, and carbon atoms, respectively.}
    \label{fig:DFT_PSB}
\end{figure*}
To gain microscopic insight into the physical origin of the dipole rotation, the experimental results are complemented by 0~K first-principles calculations~\cite{Cholsuk2023Comprehensive} on two representative defects spanning weak and strong electron-phonon coupling regimes. The first, $C_{\text{B}}C_{\text{N}}C_{\text{B}}C_{\text{N}}$, has a Huang-Rhys factor $\text{HR} = 2.71$ and configuration coordinate displacement $\Delta Q = 0.42~\mathrm{amu}^{1/2}\,\mathrm{\AA}$, representing the weak-coupling limit. The second, $C_{\text{N}}V_{\text{N}}$, has $\text{HR} = 5.96$ and $\Delta Q = 0.87~\mathrm{amu}^{1/2}\,\mathrm{\AA}$, representing the strong-coupling limit. These defects are selected to demonstrate how vibronic coupling strength governs the phonon-induced reorientation of the transition dipole, rather than to assign the experimental defect identity. Given the multi-layer nature of the drop-cast flakes, the defects are modeled in the bulk limit (Figure~\ref{fig:DFT_PSB}), with monolayer calculations provided to assess the influence of layer thickness (Supplementary Section~S3).
    
As established in the vibronic framework (Figure~\ref{figure1}a), optical emission can proceed via the ZPL or phonon-assisted pathways. For the ZPL, the defect geometries were optimized in both ground- and excited-state configurations, with the transition dipole orientation extracted from the corresponding wavefunctions. While the experimental spectra resolve the continuous acoustic manifold and discrete OPSBs, the DFT framework identifies the individual optical phonon modes that contribute to the PSB. Within this framework, Stokes phonon-sideband emission proceeds from the zero-point vibrational level of the excited state, and the transition dipole is evaluated between the fixed excited-state wavefunction and the ground-state wavefunction at each phonon-displaced geometry, where the displacement amplitude $\Delta Q_k$ is mode-dependent and reflects the projection of the excited-to-ground-state structural difference onto each phonon mode~\cite{Alkauskas2014}. To model the OPSB, we identify the vibrational modes among the $3N$ degrees of freedom and construct displaced geometries from the phonon eigenvectors. The polarization orientation is then evaluated along the phonon-assisted pathway using the excited-state wavefunctions mapped onto these displaced configurations. See Methods and Supplementary Section S3 for further details.

Figure~\ref{fig:DFT_PSB} illustrates the calculated polarization rotation within the PSB, resolved by specific optical phonon modes. For the $C_\text{B}C_\text{N}C_\text{B}C_\text{N}$ defect (Figs.~\ref{fig:DFT_PSB}a--c), the polarization deviation from the ZPL is modest, reaching a maximum of $2.7^{\circ}$. In contrast, the $C_{\text{N}}V_{\text{N}}$ defect exhibits a substantially larger response, with deviations up to $\sim 10^{\circ}$ relative to the ZPL, a factor of approximately $3.7$ larger than $C_{\text{B}}C_{\text{N}}C_{\text{B}}C_{\text{N}}$. This exceeds the ratio of their configuration coordinate displacements ($\Delta Q$ ratio $\approx 2.1$), indicating that the electronic wavefunction of $C_{\text{N}}V_{\text{N}}$ is intrinsically more sensitive to phonon-induced nuclear motion. Notably, for all configurations and modes considered, the transition dipole remains strictly in-plane, yielding a unity linear polarization visibility.

A critical point of discussion arises when comparing the mode-resolved DFT scatter to the monotonic experimental rotation. In the idealized, unstrained supercell used for DFT, individual phonon modes scatter the dipole orientation in both positive and negative directions due to the local inversion symmetry. However, the experimental measurement captures a unidirectional, continuous $40^{\circ}$ sweep. This discrepancy suggests that the local static strain inherently present in hBN flakes breaks the pristine symmetry of the environment, defining a preferred axis for the vibronic displacement. Consequently, the experiment probes the cumulative, strain-biased projection of the strongly coupled modes, leading to the observed unidirectional rotation. Consistent with our experimental findings, these results demonstrate that phonon-assisted emission induces a rotation of the polarization axis, where defects with stronger electron–phonon coupling (larger Huang–Rhys factor) exhibit larger angular deviations.

Microscopically, this rotation originates from phonon-induced atomic displacements that modify the electronic wavefunctions and thus the transition dipole moment. Stronger vibronic coupling corresponds to larger mode-projected displacements and greater sensitivity of the electronic wavefunction to phonon-induced atomic motion, both of which enhance the reorientation of the transition dipole. The difference between the calculated $\sim10^{\circ}$ optical phonon contribution and the $40^{\circ}$ room-temperature observation reflects the 0~K scope of the Franck-Condon framework, which captures Stokes emission from optical phonon modes but does not include the thermally activated acoustic phonon bath that dominates at room temperature. In 2D lattices like hBN, the acoustic phonon bath is expected to play a dominant role given the high density of states of the out-of-plane acoustic (ZA) modes at low energies and their sensitivity to local strain~\cite{Hoese2020}. The suppression of the rotation at $6$~K, where the acoustic phonon population is negligible, confirms that the 0~K framework correctly describes this thermal regime and provides direct experimental validation of the calculations, even though a quantitative first-principles account of acoustic contributions remains an important direction for future theoretical work. This nuclear-coordinate sensitivity is further evidenced by the larger angular deviation  on the anti-Stokes side of the emission, where transitions from thermally populated higher vibrational levels sample a broader range of nuclear displacements ($Q$) compared to the zero-point level, naturally amplifying the coordinate-dependency of the dipole moment $\boldsymbol{\mu}(Q)$ and leading to the observed enhancement in angular deviation. 

The direct correspondence between the spectral rotation observed in our experiments and the coordinate-dependent dipole orientation revealed by DFT suggests a unified vibronic mechanism consistent with the Herzberg-Teller framework~\cite{Herzberg.Teller.1933}, in which vibronic coupling produces a transition dipole that depends explicitly on the nuclear configuration, going beyond the Condon approximation. In the standard Condon approximation, the transition dipole is treated as a constant ($\boldsymbol{\mu}$) at the equilibrium geometry ($Q_0$). However, our findings reveal a significant departure from this conventional static dipole framework, as the dipole moment $\boldsymbol{\mu}$ is intrinsically coupled to the nuclear configuration $Q$. This coordinate-dependency, recently highlighted as a critical factor in the optical signatures of molecular systems~\cite{Kundu.Makri.2022} and solid-state defects~\cite{Quan.Duan.2025}, is described by a first-order expansion of the transition dipole moment:
\begin{equation}
\boldsymbol{\mu}(Q) \approx \boldsymbol{\mu}(Q_0) + \sum_k \left( \frac{\partial \boldsymbol{\mu}}{\partial Q_k} \right) Q_k
\end{equation}
where $Q_k$ represents the normal coordinates of the $k$-th phonon mode. The gradient term $(\partial\boldsymbol{\mu}/\partial Q_k)$ is precisely the non-Condon correction to the transition dipole moment, capturing the intrinsic sensitivity of the electronic wavefunction to displacement along mode $k$, while $Q_k$ is mode-dependent and reflects the projection of the excited-to-ground-state structural relaxation onto each phonon mode. If the gradient is not co-linear with $\boldsymbol{\mu}(Q_0)$, the transition dipole orientation undergoes a spatial reorientation as the system populates higher vibrational states. Since the $Q_k$ values are mode-dependent and in general not negligibly small, higher-order contributions may also be present, but the first-order term captures the essential physics of the coordinate-dependent rotation demonstrated here. While the DFT calculations provide discrete snapshots of this sensitivity by comparing polarization fluctuations for two distinct defect environments with their corresponding configuration, the experimental energy-resolved polarimetry probes the continuous evolution of $\boldsymbol{\mu}(Q)$ throughout the vibronic manifold~\cite{ Myers.Moerner.1994,Zirkelbach.Sandoghdar.2022}. As the emission shifts from the zero-phonon line into the phonon sidebands, the system populates higher vibrational states that sample a broader range of nuclear displacements~\cite{Qian.Chen.2020}, leading to the systematic $40^{\circ}$ rotation observed at room temperature. This interpretation is further supported by the suppression of the rotation at $6~\text{K}$, where the system remains in the ground vibrational state and the dipole is restricted to its equilibrium orientation.

\section{Conclusion}
In summary, single hBN quantum emitters exhibit a continuous, energy-resolved vectorial reorientation of the emission dipole reaching up to $40^{\circ}$ across the room-temperature vibronic manifold, demonstrating that the Condon approximation fundamentally breaks down and the transition dipole is a coordinate-dependent quantity. The suppression of this rotation at cryogenic temperatures identifies thermally activated lattice vibrations, specifically the acoustic phonon continuum, as the primary driver of the reorientation rather than a static structural property of the defect. First-principles calculations corroborate these findings, revealing that phonon-induced atomic displacements perturb the electronic wavefunctions and drive a transition dipole whose orientation depends on the nuclear coordinate. This sensitivity suggests that energy-resolved polarimetry can serve as a quantitative all-optical probe of the local phonon environment and vibronic coupling strength, capturing information that spectrally integrated measurements cannot access. Beyond identifying a fundamental limit for polarization fidelity in solid-state quantum networks, these results establish that emission polarization in atomically thin materials is a richer physical observable than static models predict.

The present observations connect solid-state quantum emitters to a broader vibronic landscape previously explored in isolated organic molecules under extreme cryogenic and vacuum conditions via scanning-tunneling-microscope-induced luminescence, where vibronic contributions to the transition dipole orientation have been resolved at sub-molecular resolution~\cite{Kong.Dong.2021, 
Vasilev.Schull.2024}. Demonstrating this in an optically-addressed solid-state host at room temperature extends this physics into a technologically accessible regime, where individual defect environments and coupling strengths can be systematically characterized. Integrating hBN emitters into surface acoustic wave devices or phononic crystal cavities offers a route toward active, phonon-engineered modulation of the emission polarization~\cite{Moores2018}. On the theoretical side, the identified dominance of acoustic phonon contributions motivates finite-temperature first-principles approaches that capture non-Condon fluctuations~\cite{Wiethorn.Montoya-Castillo.2023, Allan.Zuehlsdorff.2025}, representing a key direction for predictive modeling of non-Condon effects in solid-state quantum emitter systems.
\section{Methods}
\subsection{Sample and Experimental Setup}
Multilayer hBN flakes were obtained in solution form commercially from Graphene Supermarket. A droplet of approximately $10$~$\mu$L of solution was drop-cast onto a Si/SiO$_2$ substrate. Optical investigations were performed using two custom-built confocal microscopy systems designed for different thermal regimes. For both configurations, the excitation source was a 637 nm pulsed laser (Pilas, Advanced Laser Diode Systems, 80~ps pulse duration), with the incident polarization controlled by a half-wave plate (HWP) to maximize the PL intensity. The room-temperature measurements were conducted using a high-numerical-aperture objective lens (0.90 NA, 100× M Plan Apo HR, Mitutoyo). Conversely, the temperature-dependent studies shown in Figure~\ref{figure3} were performed between 6~K and 300~K using a custom closed-cycle cryostat (NanoMagnetics Instruments), where the sample was addressed by a low-temperature compatible objective (0.82 NA, LT-APO VIS, NanoMagnetics Instruments) inside the vacuum chamber. In both setups, the collected PL was spatially filtered and passed through a notch filter to suppress the laser light. The signal was analyzed using a spectrometer with 0.03 nm resolution (Shamrock 750, Andor) equipped with a charge-coupled device (CCD) camera (Newton, Andor). Polarization analysis was performed using either a quarter-wave plate (QWP) followed by a polarizing beam splitter (PBS) to measure the full Stokes parameters or a rotating HWP for linear polarization detection. For photon-correlation ($g^{(2)}(\tau)$) measurements, the signal was spectrally isolated using bandpass and longpass filters and detected by two single-photon avalanche photodiodes (SPCM-AQRH, Excelitas). Photon arrival times were recorded using a time-tagger unit (quTAG, Qutools).
\subsection{DFT Calculations}
All first-principles calculations were performed using the Vienna \textit{Ab initio} Simulation Package (VASP) \cite{vasp1,vasp2} with the projector augmented-wave (PAW) method \cite{paw,paw2}. We employed the screened hybrid functional HSE with a Hartree–Fock exchange mixing parameter $\alpha$=0.32, yielding a band gap of 6.09~eV for bulk AA'-stacked hBN, in excellent agreement with experiment \cite{10.1038/s41467-019-10610-5}. Van der Waals interactions were included via the D3 correction \cite{10.1063/1.3382344}. 
Defect structures were obtained from the hBN database \cite{cholsuk_advancing_2025,cholsuk_hbn_2024} and were originally computed in a \(6\times6\times2\) supercell (288 atoms) using \(\Gamma\)-point sampling and a 500~eV plane-wave cutoff, with structural relaxations converged to residual forces below \(10^{-2}\)~eV/\AA\ and total-energy changes below \(10^{-4}\)~eV. Excited-state configurations were obtained using the $\Delta$SCF approach \cite{10.1103/RevModPhys.61.689}. See Supplementary S3 for details of the polarization analysis.

\section{acknowledgement}
This work was supported by the QuantERA II Programme that has received funding from the EU Horizon 2020 research and innovation programme under GA No.~101017733 (Comphort), Scientific and Technological Research Council of Turkey (T\"{U}B\.{I}TAK) under GA Nos.~124N110 and 124N115. This research is part of the Munich Quantum Valley, which is supported by the Bavarian state government with funds from the Hightech Agenda Bayern Plus. This work was funded by the Deutsche Forschungsgemeinschaft (DFG, German Research Foundation) under Germany’s Excellence Strategy - EXC-2111-390814868 (MCQST) and as part of the CRC 1375 NOA project C2 (398816777). The authors acknowledge support from the Federal Ministry of Research, Space and Technology (BMFTR) under grant number 13N16292 (ATOMIQS). The authors gratefully acknowledge the Gauss Centre for Supercomputing e.V.\ (www.gauss-centre.eu) for funding this project by providing computing time on the GCS Supercomputer SuperMUC-NG at Leibniz Supercomputing Centre (www.lrz.de) and on its Linux-Cluster.

During the preparation of this work, the authors used AI-based language tools to improve the readability, clarity, and grammatical accuracy of the manuscript. The authors take full responsibility for the content of the published article.

\bibliographystyle{apsrev4-1}
\bibliography{references}

@article{Kianinia.Aharonovich.2022, 
year = {2022}, 
title = {{Quantum emitters in 2D materials: Emitter engineering, photophysics, and integration in photonic nanostructures}}, 
author = {Kianinia, Mehran and Xu, Zai-Quan and Toth, Milos and Aharonovich, Igor}, 
journal = {Applied Physics Reviews}, 
doi = {10.1063/5.0072091}, 
pages = {011306}, 
number = {1}, 
volume = {9}
}

@article{Esmann.Anton-Solanas.2024,
author = {Esmann, Martin and Wein, Stephen C. and Antón-Solanas, Carlos},
title = {Solid-State Single-Photon Sources: Recent Advances for Novel Quantum Materials},
journal = {Advanced Functional Materials},
volume = {34},
number = {30},
pages = {2315936},
year = {2024}
}

@article{Cakan2025,
  author    = {{\c{C}}akan, Asl{\i} and Cholsuk, Chanaprom and Gale, Angus and Kianinia, Mehran and Pa{\c{c}}al, Serkan and Ates, Serkan and Aharonovich, Igor and Toth, Milos and Vogl, Tobias},
  title     = {Quantum Optics Applications of Hexagonal Boron Nitride Defects},
  journal   = {Advanced Optical Materials},
  year      = {2025},
  volume    = {13},
  number    = {7},
  pages     = {2402508},
  doi       = {10.1002/adom.202402508},
  url       = {https://doi.org/10.1002/adom.202402508}
}

@article{Jungwirth2016,
  author    = {Jungwirth, Nicholas R. and Calderon, Brian and Ji, Yanxin and Spencer, Michael G. and Flatt{\'e}, Michael E. and Fuchs, Gregory D.},
  title     = {Temperature Dependence of Wavelength Selectable Zero-Phonon Emission from Single Defects in Hexagonal Boron Nitride},
  journal   = {Nano Letters},
  year      = {2016},
  volume    = {16},
  number    = {10},
  pages     = {6052--6057},
  publisher = {American Chemical Society},
  doi       = {10.1021/acs.nanolett.6b01987},
  issn      = {1530-6984}
}

@article{Jungwirth2017,
  title = {Optical Absorption and Emission Mechanisms of Single Defects in Hexagonal Boron Nitride},
  author = {Jungwirth, Nicholas R. and Fuchs, Gregory D.},
  journal = {Phys. Rev. Lett.},
  volume = {119},
  issue = {5},
  pages = {057401},
  numpages = {6},
  year = {2017},
  month = {Jul},
  publisher = {American Physical Society},
  doi = {10.1103/PhysRevLett.119.057401},
  url = {https://link.aps.org/doi/10.1103/PhysRevLett.119.057401}
}

@article{Tran2016,
  author    = {Toan Trong Tran and Kerem Bray and Michael J. Ford and Milos Toth and Igor Aharonovich},
  title     = {Quantum emission from hexagonal boron nitride monolayers},
  journal   = {Nature Nanotechnology},
  year      = {2016},
  volume    = {11},
  number    = {1},
  pages     = {37--41},
  doi       = {10.1038/nnano.2015.242},
  url       = {https://doi.org/10.1038/nnano.2015.242},
  issn      = {1748-3395}
}

@article{White2021,
author = {Simon White and Connor Stewart and Alexander S. Solntsev and Chi Li and Milos Toth and Mehran Kianinia and Igor Aharonovich},
journal = {Optica},
number = {9},
pages = {1153--1158},
publisher = {Optica Publishing Group},
title = {Phonon dephasing and spectral diffusion of quantum emitters in hexagonal boron nitride},
volume = {8},
month = {Sep},
year = {2021},
url = {https://opg.optica.org/optica/abstract.cfm?URI=optica-8-9-1153},
doi = {10.1364/OPTICA.431262}
}

@article{Schaefer2007,
    author = {Schaefer, Beth and Collett, Edward and Smyth, Robert and Barrett, Daniel and Fraher, Beth},
    title = {Measuring the Stokes polarization parameters},
    journal = {American Journal of Physics},
    volume = {75},
    number = {2},
    pages = {163-168},
    year = {2007},
    month = {02},
    issn = {0002-9505},
    doi = {10.1119/1.2386162},
    url = {https://doi.org/10.1119/1.2386162}
}

@article{Samaner2022,
author = {{\c C}a{\u g}lar Samaner and Serkan Pa{\c c}al and G{\"o}rkem Mutlu and K{\i}van{\c c} Uyan{\i}k and Serkan Ate{\c s}},
title = {Free-Space Quantum Key Distribution with Single Photons from Defects in Hexagonal Boron Nitride},
journal = {Advanced Quantum Technologies},
volume = {5},
number = {9},
pages = {2200059},
doi = {https://doi.org/10.1002/qute.202200059},
url = {https://advanced.onlinelibrary.wiley.com/doi/abs/10.1002/qute.202200059},
year = {2022}
}

@article{Samaner2025,
  author    = {Samaner, {\c{C}}a{\u{g}}lar and Ates, Serkan},
  title     = {Time-Resolved Stokes Polarization Analysis of Single Photon Emitters in Hexagonal Boron Nitride},
  journal   = {ACS Photonics},
  year      = {2025},
  volume    = {12},
  number    = {9},
  pages     = {5042--5049},
  doi       = {10.1021/acsphotonics.5c00988},
  url       = {https://doi.org/10.1021/acsphotonics.5c00988},
  publisher = {American Chemical Society}
}

@article{Juboori2023,
author = {Al-Juboori, Ali and Zeng, Helen Zhi Jie and Nguyen, Minh Anh Phan and Ai, Xiaoyu and Laucht, Arne and Solntsev, Alexander and Toth, Milos and Malaney, Robert and Aharonovich, Igor},
title = {Quantum Key Distribution Using a Quantum Emitter in Hexagonal Boron Nitride},
journal = {Advanced Quantum Technologies},
volume = {6},
number = {9},
pages = {2300038},
doi = {https://doi.org/10.1002/qute.202300038},
year = {2023}
}

@article{Rizzato2023,
  author = {Roberto Rizzato and Martin Schalk and Stephan Mohr and Jens C. Hermann and Joachim P. Leibold and Fleming Bruckmaier and Giovanna Salvitti and Chenjiang Qian and Peirui Ji and Georgy V. Astakhov and Ulrich Kentsch and Manfred Helm and Andreas V. Stier and Jonathan J. Finley and Dominik B. Bucher},
  title = {Extending the coherence of spin defects in hBN enables advanced qubit control and quantum sensing},
  journal   = {Nature Communications},
  year      = {2023},
  volume    = {14},
  number    = {1},
  pages     = {5089},
  doi       = {10.1038/s41467-023-40473-w},
  url       = {https://doi.org/10.1038/s41467-023-40473-w},
  issn      = {2041-1723}
}

@article{Sortino2024,
  author    = {Luca Sortino and Angus Gale and Lucca K{\"u}hner and Chi Li and Jonas Biechteler and Fedja J. Wendisch and Mehran Kianinia and Haoran Ren and Milos Toth and Stefan A. Maier and Igor Aharonovich and Andreas Tittl},
  title     = {Optically addressable spin defects coupled to bound states in the continuum metasurfaces},
  journal   = {Nature Communications},
  year      = {2024},
  volume    = {15},
  number    = {1},
  pages     = {2008},
  doi       = {10.1038/s41467-024-46272-1},
  url       = {https://doi.org/10.1038/s41467-024-46272-1},
  issn      = {2041-1723}
}

@article{Nateeboon2024,
    author = {Nateeboon, Takla and Cholsuk, Chanaprom and Vogl, Tobias and Suwanna, Sujin},
    title = {Modeling the performance and bandwidth of single-atom adiabatic quantum memories},
    journal = {APL Quantum},
    volume = {1},
    number = {2},
    pages = {026107},
    year = {2024},
    month = {04},
    issn = {2835-0103},
    doi = {10.1063/5.0188597},
    url = {https://doi.org/10.1063/5.0188597}
}

@article{Kumar2024,
  author    = {Anand Kumar and {\c C}a{\u g}lar Samaner and Chanaprom Cholsuk and Tjorben Matthes and Serkan Pa{\c c}al and Ya{\u g}{\i}z Oyun and Ashkan Zand and Robert J. Chapman and Gr{\'e}goire Saerens and Rachel Grange and Sujin Suwanna and Serkan Ate{\c s} and Tobias Vogl},
  title     = {Polarization Dynamics of Solid-State Quantum Emitters},
  journal   = {ACS Nano},
  year      = {2024},
  volume    = {18},
  number    = {7},
  pages     = {5270--5281},
  doi       = {10.1021/acsnano.3c08940},
  url       = {https://doi.org/10.1021/acsnano.3c08940},
  publisher = {American Chemical Society},
  issn      = {1936-0851}
}

@article{Becher2019,
title = {New insights into nonclassical light emission from defects in multi-layer hexagonal boron nitride},
author = {Alexander Bommer and Christoph Becher},
pages = {2041--2048},
volume = {8},
number = {11},
journal = {Nanophotonics},
doi = {doi:10.1515/nanoph-2019-0123},
year = {2019},
}

@article{Wigger2019,
doi = {10.1088/2053-1583/ab1188},
year = {2019},
month = {apr},
publisher = {IOP Publishing},
volume = {6},
number = {3},
pages = {035006},
author = {Wigger, Daniel and Schmidt, Robert and Del Pozo-Zamudio, Osvaldo and PreuB, Johann A and Tonndorf, Philipp and Schneider, Robert and Steeger, Paul and Kern, Johannes and Khodaei, Yashar and Sperling, Jaroslaw and de Vasconcellos, Steffen Michaelis and Bratschitsch, Rudolf and Kuhn, Tilmann},
title = {Phonon-assisted emission and absorption of individual color centers in hexagonal boron nitride},
journal = {2D Materials},
}

@article{Hoese2020,
author = {Michael Hoese  and Prithvi Reddy  and Andreas Dietrich  and Michael K. Koch  and Konstantin G. Fehler  and Marcus W. Doherty  and Alexander Kubanek },
title = {Mechanical decoupling of quantum emitters in hexagonal boron nitride from low-energy phonon modes},
journal = {Science Advances},
volume = {6},
number = {40},
pages = {eaba6038},
year = {2020},
doi = {10.1126/sciadv.aba6038},
}

@article{Nikolay2019,
author = {Niko Nikolay and Noah Mendelson and Ersan \"{O}zelci and Bernd Sontheimer and Florian B\"{o}hm and G\"{u}nter Kewes and Milos Toth and Igor Aharonovich and Oliver Benson},
journal = {Optica},
number = {8},
pages = {1084--1088},
publisher = {Optica Publishing Group},
title = {Direct measurement of quantum efficiency of single-photon emitters in hexagonal boron nitride},
volume = {6},
month = {Aug},
year = {2019},
url = {https://opg.optica.org/optica/abstract.cfm?URI=optica-6-8-1084},
doi = {10.1364/OPTICA.6.001084}
}

@article{Dietrich2018,
  title = {Observation of Fourier transform limited lines in hexagonal boron nitride},
  author = {Dietrich, A. and B\"urk, M. and Steiger, E. S. and Antoniuk, L. and Tran, T. T. and Nguyen, M. and Aharonovich, I. and Jelezko, F. and Kubanek, A.},
  journal = {Phys. Rev. B},
  volume = {98},
  issue = {8},
  pages = {081414},
  numpages = {5},
  year = {2018},
  month = {Aug},
  publisher = {American Physical Society},
  doi = {10.1103/PhysRevB.98.081414},
  url = {https://link.aps.org/doi/10.1103/PhysRevB.98.081414}
}

@article{Grosso2020,
author = {Grosso, Gabriele and Moon, Hyowon and Ciccarino, Christopher J. and Flick, Johannes and Mendelson, Noah and Mennel, Lukas and Toth, Milos and Aharonovich, Igor and Narang, Prineha and Englund, Dirk R.},
title = {Low-Temperature Electron–Phonon Interaction of Quantum Emitters in Hexagonal Boron Nitride},
journal = {ACS Photonics},
volume = {7},
number = {6},
pages = {1410-1417},
year = {2020},
doi = {10.1021/acsphotonics.9b01789}
}

@article{Cusco2016,
  title = {Temperature dependence of Raman-active phonons and anharmonic interactions in layered hexagonal BN},
  author = {Cusc\'o, Ramon and Gil, Bernard and Cassabois, Guillaume and Art\'us, Luis},
  journal = {Phys. Rev. B},
  volume = {94},
  issue = {15},
  pages = {155435},
  numpages = {11},
  year = {2016},
  month = {Oct},
  publisher = {American Physical Society},
  doi = {10.1103/PhysRevB.94.155435},
  url = {https://link.aps.org/doi/10.1103/PhysRevB.94.155435}
}

@article{Exarhos2017,
  author    = {Exarhos, Annemarie L. and Hopper, David A. and Grote, Richard R. and Alkauskas, Audrius and Bassett, Lee C.},
  title     = {Optical Signatures of Quantum Emitters in Suspended Hexagonal Boron Nitride},
  journal   = {ACS Nano},
  year      = {2017},
  volume    = {11},
  number    = {3},
  pages     = {3328--3336},
  publisher = {American Chemical Society},
  doi       = {10.1021/acsnano.7b00665},
  url       = {https://doi.org/10.1021/acsnano.7b00665},
  issn      = {1936-0851}
}

@article{Khatri2019,
  title = {Phonon sidebands of color centers in hexagonal boron nitride},
  author = {Khatri, P. and Luxmoore, I. J. and Ramsay, A. J.},
  journal = {Phys. Rev. B},
  volume = {100},
  issue = {12},
  pages = {125305},
  numpages = {7},
  year = {2019},
  month = {Sep},
  publisher = {American Physical Society},
  doi = {10.1103/PhysRevB.100.125305},
  url = {https://link.aps.org/doi/10.1103/PhysRevB.100.125305}
}

@article{Jha2022,
doi = {10.1088/1361-6528/ac2b71},
url = {https://dx.doi.org/10.1088/1361-6528/ac2b71},
year = {2021},
month = {oct},
publisher = {IOP Publishing},
volume = {33},
number = {1},
pages = {015001},
author = {Jha, Pankaj K and Akbari, Hamidreza and Kim, Yonghwi and Biswas, Souvik and Atwater, Harry A},
title = {Nanoscale axial position and orientation measurement of hexagonal boron nitride quantum emitters using a tunable nanophotonic environment},
journal = {Nanotechnology}
}

@article{Krummheuer2002,
  title = {Theory of pure dephasing and the resulting absorption line shape in semiconductor quantum dots},
  author = {Krummheuer, B. and Axt, V. M. and Kuhn, T.},
  journal = {Phys. Rev. B},
  volume = {65},
  issue = {19},
  pages = {195313},
  numpages = {12},
  year = {2002},
  month = {May},
  publisher = {American Physical Society},
  doi = {10.1103/PhysRevB.65.195313},
  url = {https://link.aps.org/doi/10.1103/PhysRevB.65.195313}
}

@article{Kozlov2018,
  title = {Hidden polarization of unpolarized light},
  author = {Kozlov, G. G. and Ryzhov, I. I. and Tzimis, A. and Hatzopoulos, Z. and Savvidis, P. G. and Kavokin, A. V. and Bayer, M. and Zapasskii, V. S.},
  journal = {Phys. Rev. A},
  volume = {98},
  issue = {4},
  pages = {043810},
  numpages = {6},
  year = {2018},
  month = {Oct},
  publisher = {American Physical Society},
  doi = {10.1103/PhysRevA.98.043810},
  url = {https://link.aps.org/doi/10.1103/PhysRevA.98.043810}
}

@article{vasp1,
title = "Efficiency of Ab-Initio Total Energy Calculations for Metals and Semiconductors using a Plane-Wave Basis Set",
journal = "Comput. Mater. Sci.",
volume = "6",
number = "1",
pages = "15 - 50",
year = "1996",
issn = "0927-0256",
doi = "https://doi.org/10.1016/0927-0256(96)00008-0",
author = "G. Kresse and J. Furthmüller"
}

@article{vasp2,
  title = {Efficient Iterative Schemes for Ab Initio Total-Energy Calculations using a Plane-Wave Basis Set},
  author = {Kresse, G. and Furthm\"uller, J.},
  journal = {Phys. Rev. B},
  volume = {54},
  issue = {16},
  pages = {11169},
  numpages = {0},
  year = {1996},
  publisher = {American Physical Society},
  doi = {https://doi.org/10.1103/PhysRevB.54.11169},
}

@article{paw,
  title = {Projector Augmented-Wave Method},
  author = {Bl\"ochl, P. E.},
  journal = {Phys. Rev. B},
  volume = {50},
  issue = {24},
  pages = {17953},
  numpages = {0},
  year = {1994},
  month = {Dec},
  publisher = {American Physical Society},
  doi = {https://doi.org/10.1103/PhysRevB.50.17953}
}

@article{paw2,
  title = {From Ultrasoft Pseudopotentials to the Projector Augmented-Wave Method},
  author = {Kresse, G. and Joubert, D.},
  journal = {Phys. Rev. B},
  volume = {59},
  issue = {3},
  pages = {1758},
  numpages = {0},
  year = {1999},
  month = {Jan},
  publisher = {American Physical Society},
  doi = {https://doi.org/10.1103/PhysRevB.59.1758}
}

@Article{10.1038/s41467-019-10610-5,
author={Elias, C. and Valvin, P. and Pelini, T. and Summerfield, A. and Mellor, C. J. and Cheng, T. S. and Eaves, L. and Foxon, C. T. and Beton, P. H. and Novikov, S. V. and Gil, B. and Cassabois, G.},
title={Direct band-gap crossover in epitaxial monolayer boron nitride},
journal={Nat. Commun.},
year={2019},
volume={10},
number={1},
pages={2639},
url={https://doi.org/10.1038/s41467-019-10610-5},
doi = {10.1038/s41467-019-10610-5}
}

@article{10.1063/1.3382344,
	author = {Grimme, Stefan and Antony, Jens and Ehrlich, Stephan and Krieg, Helge},
	doi = {10.1063/1.3382344},
	issn = {0021-9606},
	journal = {The Journal of Chemical Physics},
	month = apr,
	number = {15},
	pages = {154104},
	title = {A consistent and accurate ab initio parametrization of density functional dispersion correction ({DFT}-{D}) for the 94 elements {H}-{Pu}},
	url = {https://doi.org/10.1063/1.3382344},
	urldate = {2025-07-03},
	volume = {132},
	year = {2010}}

@article{10.1103/RevModPhys.61.689,
   author = {R. O. Jones and O. Gunnarsson},
   doi = {10.1103/RevModPhys.61.689},
   issn = {0034-6861},
   issue = {3},
   journal = {Rev. Mod. Phys.},
   month = {7},
   pages = {689},
   title = {The Density Functional Formalism, Its Applications and Prospects},
   volume = {61},
   url = {https://link.aps.org/doi/10.1103/RevModPhys.61.689},
   year = {1989},
}

@article{Alkauskas2014,
	author = {Alkauskas, Audrius and Buckley, Bob B and Awschalom, David D and Van De Walle, Chris G},
	copyright = {http://iopscience.iop.org/info/page/text-and-data-mining},
	doi = {10.1088/1367-2630/16/7/073026},
	issn = {1367-2630},
	journal = {New J. Phys.},
	month = jul,
	number = {7},
	pages = {073026},
	title = {First-principles theory of the luminescence lineshape for the triplet transition in diamond {NV} centres},
	url = {https://iopscience.iop.org/article/10.1088/1367-2630/16/7/073026},
	urldate = {2025-04-02},
	volume = {16},
	year = {2014}}

@article{cholsuk_identifying_2024,
	author = {Cholsuk, Chanaprom and Cakan, Asli and Suwanna, Sujin and Vogl, Tobias},
	doi = {10.1002/adom.202302760},
	issn = {2195-1071, 2195-1071},
	journal = {Advanced Optical Materials},
	language = {en},
	month = may,
	number = {13},
	pages = {2302760},
	title = {Identifying electronic transitions of defects in hexagonal boron nitride for quantum memories},
	url = {https://doi.org/10.1002/adom.202302760},
	urldate = {2024-06-17},
	volume = {12},
	year = {2024}}

@article{cholsuk_advancing_2025,
	author = {Cholsuk, Chanaprom and Suwanna, Sujin and Vogl, Tobias},
	doi = {10.1039/D5TC02805A},
	issn = {2050-7526, 2050-7534},
	journal = {J. Mater. Chem. C},
	language = {en},
	number = {43},
	pages = {21826--21837},
	title = {Advancing the {hBN} defects database through photophysical characterization of bulk {hBN}},
	url = {https://xlink.rsc.org/?DOI=D5TC02805A},
	urldate = {2025-12-22},
	volume = {13},
	year = {2025}}

@article{cholsuk_hbn_2024,
	author = {Cholsuk, Chanaprom and Zand, Ashkan and {\c C}akan, Aslı and Vogl, Tobias},
	copyright = {https://creativecommons.org/licenses/by/4.0/},
	doi = {10.1021/acs.jpcc.4c03404},
	issn = {1932-7447, 1932-7455},
	journal = {J. Phys. Chem. C},
	language = {en},
	month = aug,
	number = {30},
	pages = {12716--12725},
	shorttitle = {The {hBN} {Defects} {Database}},
	title = {The {hBN} {Defects} {Database}: {A} {Theoretical} {Compilation} of {Color} {Centers} in {Hexagonal} {Boron} {Nitride}},
	url = {https://pubs.acs.org/doi/10.1021/acs.jpcc.4c03404},
	urldate = {2024-08-01},
	volume = {128},
	year = {2024}}

@article{Moores2018,
  title = {Cavity Quantum Acoustic Device in the Multimode Strong Coupling Regime},
  author = {Moores, Bradley A. and Sletten, Lucas R. and Viennot, Jeremie J. and Lehnert, K. W.},
  journal = {Phys. Rev. Lett.},
  volume = {120},
  issue = {22},
  pages = {227701},
  numpages = {5},
  year = {2018},
  month = {May},
  publisher = {American Physical Society},
  doi = {10.1103/PhysRevLett.120.227701},
  url = {https://link.aps.org/doi/10.1103/PhysRevLett.120.227701}
}

@article{Cholsuk2023Comprehensive,
  author = {Cholsuk, Chanaprom and Suwanna, Sujin and Vogl, Tobias},
  title = {Comprehensive Scheme for Identifying Defects in Solid-State Quantum Systems},
  journal = {The Journal of Physical Chemistry Letters},
  year = {2023},
  volume = {14},
  number = {29},
  pages = {6564--6571},
  doi = {10.1021/acs.jpclett.3c01475},
  url = {https://doi.org/10.1021/acs.jpclett.3c01475},
  publisher = {American Chemical Society},
  month = {07}
}

@article{Gerard.Delteil.2026, 
year = {2026}, 
title = {{Resonance fluorescence and indistinguishable photons from a coherently driven B centre in hBN}}, 
author = {Gérard, Domitille and Buil, Stéphanie and Watanabe, Kenji and Taniguchi, Takashi and Hermier, Jean-Pierre and Delteil, Aymeric}, 
journal = {Nature Communications}, 
doi = {10.1038/s41467-026-68555-5}, 
pages = {1843}, 
number = {1}, 
volume = {17}
}

@article{Martinez.Solanas.2026, 
year = {2026}, 
title = {{Temporal Coherence of Single Photons Emitted by Hexagonal Boron Nitride Defects at Room Temperature}}, 
author = {Mart{\'\i}nez-Pons, Juan Vidal and Kim, Sang Kyu and Behrens, Max and Izquierdo-Molina, Alejandro and Rua, Adolfo Menendez and Pa{\c{c}}al, Serkan and Ate{\c{s}}, Serkan and Vi{\~n}a, Luis and Ant{\'o}n-Solanas, Carlos}, 
journal = {ACS Photonics}, 
issn = {2330-4022}, 
doi = {10.1021/acsphotonics.5c02227}, 
pages = {282--289}, 
number = {1}, 
volume = {13}
}

@article{Tapsin.Ates.2025, 
year = {2025}, 
title = {{Secure Quantum Key Distribution with Room-Temperature Quantum Emitter}}, 
author = {Tap{\c{s}}{\i}n, {\"O}mer S. and A{\u{g}}larc{\i}, Furkan and Pousa, Roberto G. and Oi, Daniel K. L. and G{\"u}ndo{\u{g}}an, Mustafa and Ate{\c{s}}, Serkan},
eprint = {arXiv:2501.13902}
}

@article{Kubanek.Kubanek.2022, 
year = {2022}, 
title = {{Coherent Quantum Emitters in Hexagonal Boron Nitride}}, 
author = {Kubanek, Alexander}, 
journal = {Advanced Quantum Technologies}, 
issn = {2511-9044}, 
doi = {10.1002/qute.202200009}, 
pages = {2200009}, 
number = {9}, 
volume = {5}
}

@article{Sontheimer.Benson.2017, 
year = {2017}, 
title = {{Photodynamics of quantum emitters in hexagonal boron nitride revealed by low-temperature spectroscopy}}, 
author = {Sontheimer, Bernd and Braun, Merle and Nikolay, Niko and Sadzak, Nikola and Aharonovich, Igor and Benson, Oliver}, 
journal = {Physical Review B}, 
issn = {2469-9950}, 
doi = {10.1103/physrevb.96.121202},  
pages = {121202}, 
number = {12}, 
volume = {96} 
}

@article{Fournier.Delteil.2023, 
year = {2023}, 
title = {{Two-Photon Interference from a Quantum Emitter in Hexagonal Boron Nitride}}, 
author = {Fournier, Clarisse and Roux, S{\'e}bastien and Watanabe, Kenji and Taniguchi, Takashi and Buil, St{\'e}phanie and Barjon, Julien and Hermier, Jean-Pierre and Delteil, Aymeric},
journal = {Physical Review Applied}, 
doi = {10.1103/physrevapplied.19.l041003}, 
pages = {L041003}, 
number = {4}, 
volume = {19}
}

@article{Ari.Ates2025, 
  year    = {2025}, 
  title   = {Temperature-Dependent Spectral Properties of Hexagonal Boron Nitride Color Centers}, 
  author = {Ar{\i}, Ozan and Polat, Nahit and F{\i}rat, Volkan and {\c{C}}ak{\i}r, {\"O}zg{\"u}r and Ate{\c{s}}, Serkan},
  journal = {{ACS} Photonics}, 
  issn    = {2330-4022}, 
  doi     = {10.1021/acsphotonics.4c02616}, 
  pages   = {1676--1682}, 
  number  = {3}, 
  volume  = {12}
}

@article{Herzberg.Teller.1933, 
year = {1933}, 
title = {{Schwingungsstruktur der Elektronenübergänge bei mehratomigen Molekülen}}, 
author = {Herzberg, G. and Teller, E.}, 
journal = {Zeitschrift für Physikalische Chemie}, 
issn = {0942-9352}, 
doi = {10.1515/zpch-1933-2136}, 
pages = {410--446}, 
number = {1}, 
volume = {21B}
}

@article{Small.Small.1971, 
year = {1971}, 
title = {{Herzberg–Teller Vibronic Coupling and the Duschinsky Effect}}, 
author = {Small, Gerald J}, 
journal = {The Journal of Chemical Physics}, 
issn = {0021-9606}, 
doi = {10.1063/1.1675343}, 
pages = {3300--3306}, 
number = {8}, 
volume = {54}
}

@article{Zirkelbach.Sandoghdar.2022, 
year = {2022}, 
title = {{High-resolution vibronic spectroscopy of a single molecule embedded in a crystal}}, 
author = {Zirkelbach, Johannes and Mirzaei, Masoud and Deperasińska, Irena and Kozankiewicz, Boleslaw and Gurlek, Burak and Shkarin, Alexey and Utikal, Tobias and Götzinger, Stephan and Sandoghdar, Vahid}, 
journal = {The Journal of Chemical Physics}, 
issn = {0021-9606}, 
doi = {10.1063/5.0081297}, 
pmid = {35291792}, 
eprint = {2112.04806}, 
pages = {104301}, 
number = {10}, 
volume = {156}
}

@article{Quan.Duan.2025, 
year = {2025}, 
title = {{Significance of Herzberg-Teller effects on allowed transitions of defects in solids}}, 
author = {Quan, Qianshan and Duan, Chang-Kui}, 
journal = {Physical Review B}, 
issn = {2469-9950}, 
doi = {10.1103/physrevb.111.075156}, 
pages = {075156}, 
number = {7}, 
volume = {111}
}

@article{Qian.Chen.2020, 
year = {2020}, 
title = {{Herzberg–Teller Effect on the Vibrationally Resolved Absorption Spectra of Single-Crystalline Pentacene at Finite Temperatures}}, 
author = {Qian, Yuqin and Li, Xia and Harutyunyan, Avetik R. and Chen, Gugang and Rao, Yi and Chen, Hanning}, 
journal = {The Journal of Physical Chemistry A}, 
issn = {1089-5639}, 
doi = {10.1021/acs.jpca.0c07896}, 
pages = {9156--9165}, 
number = {44}, 
volume = {124}
}

@article{Kundu.Makri.2022, 
year = {2022}, 
title = {{Franck–Condon and Herzberg–Teller Signatures in Molecular Absorption and Emission Spectra}}, 
author = {Kundu, Sohang and Roy, Partha Pratim and Fleming, Graham R. and Makri, Nancy}, 
journal = {The Journal of Physical Chemistry B}, 
issn = {1520-6106}, 
doi = {10.1021/acs.jpcb.2c00846}, 
pages = {2899--2911}, 
number = {15}, 
volume = {126}
}

@article{Myers.Moerner.1994, 
year = {1994}, 
title = {{Vibronic Spectroscopy of Individual Molecules in Solids}}, 
author = {Myers, Anne B and Tchenio, Paul and Zgierski, Marek Z and Moerner, W E}, 
journal = {The Journal of Physical Chemistry}, 
issn = {0022-3654}, 
doi = {10.1021/j100092a001}, 
pages = {10377--10390}, 
number = {41}, 
volume = {98}
}

@article{Kong.Dong.2021,
  author  = {Kong, Fan-Fang and Tian, Xiao-Jun and Zhang, Yang 
             and Yu, Yun-Jie and Jing, Shi-Hao and Zhang, Yao 
             and Tian, Guang-Jun and Luo, Yi and Yang, Jin-Long 
             and Dong, Zhen-Chao and Hou, J. G.},
  title   = {Probing intramolecular vibronic coupling through 
             vibronic-state imaging},
  journal = {Nature Communications},
  volume  = {12},
  pages   = {1280},
  year    = {2021},
  doi     = {10.1038/s41467-021-21571-z}
}

@article{Vasilev.Schull.2024,
  author  = {Vasilev, Kirill and Canola, Sofia and Scheurer, 
             Fabrice and Boeglin, Alex and Lotthammer, Felix 
             and Ch\'{e}rioux, Fr\'{e}d\'{e}ric and Neuman, Toma\v{s} 
             and Schull, Guillaume},
  title   = {Exploring the Role of Excited States' Degeneracy 
             on Vibronic Coupling with Atomic-Scale Optics},
  journal = {ACS Nano},
  volume  = {18},
  pages   = {28052--28059},
  year    = {2024},
  doi     = {10.1021/acsnano.4c07136}
}

@article{Wiethorn.Montoya-Castillo.2023, 
year = {2023}, 
title = {{Beyond the Condon limit: Condensed phase optical spectra from atomistic simulations}}, 
author = {Wiethorn, Zachary R. and Hunter, Kye E. and Zuehlsdorff, Tim J. and Montoya-Castillo, Andrés}, 
journal = {The Journal of Chemical Physics}, 
issn = {0021-9606}, 
doi = {10.1063/5.0180405},  
pages = {244114}, 
number = {24}, 
volume = {159}
}

@article{Allan.Zuehlsdorff.2025, 
year = {2025}, 
title = {{FC2DES+HT: Including Herzberg–Teller Effects in the Simulation of 2D Electronic Spectra for Harmonic Hamiltonians}}, 
author = {Allan, Lucas and Zuehlsdorff, Tim J.}, 
journal = {Journal of Chemical Theory and Computation}, 
issn = {1549-9618}, 
doi = {10.1021/acs.jctc.5c01363}, 
pages = {11137--11151}, 
number = {21}, 
volume = {21}
}
\end{document}